\documentclass[aps, prl,12pt,reprint]{revtex4-2}

\usepackage{amsmath,amsfonts,amssymb,amsthm,yhmath}  
\usepackage{graphicx}
\usepackage{soul}

\def \e{\mathbf{e}}

\def \A{\textbf{A}}
\def \B{\textbf{B}}

\def \bq{\begin{equation}}
\def \eq{\end{equation}}
\def \bqa{\begin{eqnarray}}
\def \eqa{\end{eqnarray}}

\begin{document}

\title{Comment on "Near-Field Spin Chern Number Quantized by Real-Space Topology of Optical Structures", T. Fu et al. Phys. Rev. Lett. 132, 233801 (2024)}

\author{Didier Felbacq, Emmanuel Rousseau}

\affiliation{{L2C, University of Montpellier},
	{Place Bataillon}, 
	{Montpellier},
	{34095}, 
	{France}}


	
	
\maketitle

In \cite{PRL}, the authors claim to have introduced a "real-space spin Chern number" as well as a "Spin Berry connection" and a "Spin Berry curvature". The main finding of their letter is the statement that the integral of the "Spin Berry curvature" over the surface is equal to the "Spin Chern number" which is the Euler characteristic of the surface. What the authors show is that, given a vector field tangent to a surface, there is a connection whose curvature gives the Euler characteristic when it is integrated over the surface. The point of this comment is to explain that no new invariant has been defined and that the result shown is the exact statement of the Chern-Gauss-Bonnet theorem \cite{docarmo}, in the particular case of a surface. Since the "real-space spin Chern number" is equal to the Euler characteristic, it is not a new invariant but just another name for the same thing. Moreover, the Euler number characterizes the surface and {\it not} the polarization state of the field.

More precisely, let $M$ be a compact oriented surface with no boundary and $V$ be a tangent vector field. Denoting $\Pi_x$ the projector on the tangent plane at $x \in M$, $\Pi_x dV$ defines a connection 1-form $\omega$: $\Pi_x dV=\omega V$. The Chern-Gauss-Bonnet states that $\int_M d\omega=\chi(M)$, where $\chi(M)$ is the Euler characteristic of $M$. Given any other connection 1-form $\omega'$, it holds that $\omega-\omega'$ is globally defined, so that, given that $M$ has no boundary, $\int_M d\omega=\int_M d\omega'=\chi(M)$. 

Given a field with an arbitrary state of polarization, the authors define an oriented trihedron $(\e_A,\sigma \e_B, \hat{n})$, where $\hat{n}$ is the exterior normal to the surface. 
Denoting $\e'_B=\sigma \e_B$,
we write: $\e_A=A^j \e_j$ and $\e'_B=B'^j \e_j$, where $(\e_j)$ is the canonical basis of $\mathbb{R}^3$ and $(dx^j)$ the dual basis. We denote $\omega_A$ (resp. $\omega'_{B}$) the 1-form dual to $\e_A$ (resp. $\e'_B$). 
%
Taking the differential of $\{\e_A,\e'_B\}$ and projecting them upon the tangent space leads to:
$$
(\e_A,d\e_A)=0,\, (\e'_B,d\e'_B)=0.
$$
Furthermore:
$$
d\e_A=\partial_j A^i dx^j \otimes \e_i ,\,\,d\e'_B=\partial_j B'^i dx^j \otimes \e_i.
$$
The coefficients of the connection 1-form are given by: $\omega_{AB}(\e_k)=(\e'_B,d\e_A)(\e_k)=B'^i \partial_k A^i$.
Computing the differential of $\omega_{BA}$ gives the curvature 2-form $\Omega$ and the gaussian curvature $K$ of the surface:
$$
\Omega=d\omega_{BA}=(d\e_A,d\e'_B)=(\partial_k A^i \partial_j B^i) dx^j\wedge dx^k=K \omega_A \wedge \omega'_B.
$$
That the normalized integral of $K$ is equal to the Euler characteristic of the surface is the statement of the Chern-Gauss-Bonnet theorem \cite[p. 102]{docarmo} for surfaces.

Let us now consider the connection 1-form chosen by the authors. Denoting $\e=\A+i\B$, where $\A=A \e_A$ and $\B'=B\e'_B$, it is : $\tilde{\omega}_{BA}=\Im(\e^*,d\e)=-2(\B',d\A)$. At a singularity, it holds $\|A\|=\|B\|=\sqrt{2}/2$. Therefore, the 1-form $\tilde{\omega}_{BA}-{\omega}_{BA}$ is a \textit{globally defined 1-form} on $M$, hence the curvatures both integrate to the Euler characteristic: $\int_M \tilde{\omega}_{BA}=\int_M {\omega}_{BA}=\chi(M)$.


In conclusion, it cannot be claimed that a new invariant has been defined. The result obtained by the authors is but an application of the Chern-Gauss-Bonnet theorem to a specific case. Moreover, the claim that the results extend to real space the notions of connection, curvature and Chern class is a misconception, since these notions apply to any space of parameters, whether it be of direct or reciprocal nature.

%
%
%
%
%
%
%
%

\end{document}